\documentclass[twocolumn,showpacs,preprintnumbers,amsmath,amssymb]{revtex4}

\usepackage{graphicx}
\usepackage{bm}

\newcommand\Tm{\langle\mathbf{T}\rangle}
\newcommand\rr{{\bf r}}

\begin{document}

\title{The narrow escape problem revisited}

\author{O. B\'enichou}
\affiliation{Laboratoire de Physique Th\'eorique de la Mati\`ere Condens\'ee
(UMR 7600), Universit\'e Pierre et Marie Curie, 4 Place Jussieu, 75255
Paris Cedex}

\author{R. Voituriez}
\affiliation{Laboratoire de Physique Th\'eorique de la Mati\`ere Condens\'ee
(UMR 7600), Universit\'e Pierre et Marie Curie, 4 Place Jussieu, 75255
Paris Cedex}

\date{\today}

\begin{abstract}
The time needed for a particle to exit a confining domain through a small window, called the narrow escape time (NET), is a limiting factor of various  processes, such  as some  biochemical reactions in cells. Obtaining an estimate of the mean NET for a given geometric environment is therefore a requisite step to quantify the reaction rate constant of such processes, which has raised a growing interest in the last few years. In this Letter,  we determine explicitly the scaling dependence of the  mean NET  on both the volume of the confining domain and the starting point to aperture distance. We show that this analytical approach is applicable to a very wide range of stochastic processes, including anomalous diffusion or diffusion in the presence of an external force field, which cover situations of biological relevance.
\end{abstract}

\pacs{87.10.+e,05.40.Fb,05.40.Jc}

\maketitle
The first-passage time (FPT), namely the time it takes  a random walker to reach a given target site is known to be a key quantity to quantify the dynamics of various processes of practical interest \cite{RednerBook,nature2007,nv2007}. Indeed, biochemical reactions \cite{SlutskyBiophys04,nousprotein,Eliazar2007,Kolesov2007},  foraging strategies of animals \cite{VisNature1999,nousanimaux,Benichou2006}, the spread of sexually transmitted diseases in a human social network or of viruses through the world wide web \cite{songpnas} are often controlled by first encounter events.

\begin{figure}
\centering\includegraphics[width =0.6\linewidth,clip]{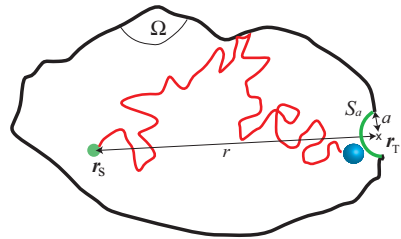}
\caption{The narrow escape problem : the particle starts from  $\rr_S$ and evolves in a domain $\Omega$ with reflecting walls, except a small aperture $S_a$ of typical radius $a$ centered at $\rr_T$.}
\label{fig1}
\end{figure}

Among first-passage processes, the case where the target is a small window on the boundary of a confining domain, defined as the narrow escape problem, has proved very recently to be of particular importance \cite{Schuss2007}. The narrow escape time (NET) gives the time needed for a random walker trapped in a confining domain with a single narrow opening to exit the domain for the first time (see fig.\ref{fig1}). The relevance of the NET is striking in cellular biology, since it gives for instance the time needed for a reactive particle  released from a specific organelle to activate a given protein on the cell membrane \cite{alberts}. Further examples are given by biochemical reactions in cellular microdomains, like dentritic spines, synapses or microvesicles to name a few \cite{alberts,Schuss2007}. These submicrometer domains often contain a small amount of particles which must first exit the domain in order to fulfill their biological function. In these examples, the NET is therefore a limiting quantity whose quantization is a first step in the modeling of the process.

An important theoretical advance has been made recently by different groups \cite{Grigoriev02,Singer06,Schuss2007}, which obtained  the leading term of the mean NET in the limit of small aperture in the case of a brownian particle. However, these different approaches lose track of the  dependence of the mean NET on the starting point. Obtaining such information is not only an important theoretical issue, but also a biologically  relevant question. As a matter of fact, biomolecules like membrane signalling proteins or  transcription factor proteins are generated at specific sites in the cell \cite{alberts}, whose localization plays an important role in the very function of the biomolecules, as underlined recently in  \cite{Kolesov2007}.  In addition, the above mentioned techniques to estimate mean NETs have been limited so far to normal brownian diffusion, whereas many experimental studies have shown that cellular transport often departs from thermal diffusion  due to the complexity of the cellular environment. In particular, crowding effects have proved to induce subdiffusive behavior  in many situations \cite{Golding2006,Platani2002}, while  the interaction of a tracer particle with molecular motors induces a  biased motion \cite{alberts}.

In this Letter we propose a general theory  which (i) provides explicitly the scaling dependence of the mean NET on both the volume of the confining domain and the source-target distance, (ii) applies to a wide range of transport processes, including anomalous diffusion, (iii) encompasses the case of transport in the presence of a force field. Our formalism partially relies on the method recently proposed  in \cite{nature2007}, and considerably broadens its field of applications.

We start by extending the theory developed in \cite{nature2007} to compute the mean FPT of a continuous random motion to a closed surface $S_a({\bf r}_T)$ of typical radius $a$ containing  ${\bf r}_T$, starting from a source point ${\bf r}_S$. We will next give explicit results when $S_a({\bf r}_T)$ is a sphere, and show that this approach allows to solve the narrow escape problem. The volume delimited by $S_a({\bf r}_T)$ will be denoted by $B_a({\bf r}_T)$.
We consider that the random walker
evolves in a bounded domain $\Omega$ of volume  $V$ of the $d$--dimensional space $\mathbb{R}^d$.
Let $P({\bf r},t|{\bf r}')$ be the propagator, i.e. the density probability to be
at  ${\bf r}$ at time $t$, starting from  ${\bf r}'$ at time $0$, which satisfies the  backward equation \cite{Kampen1992}:
\begin{equation}
\frac{{\partial}}{{\partial t}} P({\bf r},t|{\bf r'})=\Delta_{\bf r'}P({\bf r},t|{\bf r'}),\label{propa}
\end{equation}
where $ \Delta_{\bf r}$ denotes the Laplace operator and  the diffusion coefficient has been set to 1. Let  $F({\bf r}',t|{\bf r}) d S({\bf r}')$ be the probability that the first-passage time
at the infinitesimal surface $d S({\bf r}')$ located at ${\bf r}'$, starting from ${\bf r}$, is $t$. By partitioning over the first arrival time $t'$ at a surface element $dS({\bf r}')$  of the sphere $S_a({\bf r}_T)$, one obtains
\begin{equation}\label{eq2}
P({\bf r}_T,t|{\bf r}_S) = \int_{{\bf r}\in S_a({\bf r}_T)}\!\!\!\!\!\!\!\!\!\!\!\! d S({\bf r})\int_0^t\!\!\!
P({\bf r}_T,t-t'|{\bf r})F({\bf r},t'|{\bf r}_S) dt'.
\end{equation}
We next assume $a\ll V^{1/d}$, so that $ P({\bf r}_T,t|{\bf r}\in S_a({\bf r}_T))\equiv P({\bf r}_T,t|S_a({\bf r}_T))$ does not depend on ${\bf r}\in S_a({\bf r}_T)$. This condition will be fulfilled in the large $V$ limit considered in the following. Denoting $\widehat{p}(s)=\int_0^\infty p(t) e^{-st}dt$ the  Laplace transform of $p(t)$, and setting
\begin{equation}
 F(S_a({\bf r}_T),t|{\bf r}_S)=\int_{{\bf r}\in S_a({\bf r}_T)}\!\!\!\!\!\!\!\!\!\!\!\!\!\!\!\!\!\!\!dS({\bf r})F({\bf r},t|{\bf r}_S),
\end{equation}
we get from equation (\ref{eq2}):

\begin{equation}\label{laplace}
 \widehat{P}({\bf r}_T,s|{\bf r}_S)=\widehat{P}({\bf r}_T,s|S_a({\bf r}_T))\widehat{F}(S_a({\bf r}_T),s|{\bf r}_S).
\end{equation}
We now denote by $\Tm(S_a({\bf r}_T)|{\bf r}_S)$ the mean FPT  at $S_a({\bf r}_T)$, and write $\lim_{t\to\infty}P({\bf r}_T,t|{\bf r}_S)= P_{\rm stat}({\bf r}_T)$. Then one has
\begin{equation}
 \left\{
\begin{array}{ll}
 & \widehat{F}(S_a({\bf r}_T),s|{\bf r}_S) =1-s\Tm(S_a({\bf r}_T)|{\bf r}_S)+o(s)\\
 & \widehat{P}({\bf r}_T,s|{\bf r}_S)  =  P_{\rm stat}({\bf r}_T)/s+H({\bf r}_T|{\bf r}_S)+o(1)
\end{array}\right.
\end{equation} where

\begin{equation}
H({\bf r}|{\bf r}') = \int_{0}^{\infty} (P({\bf r},t|{\bf r}') - P_{\rm stat}({\bf r}))
dt.\label{pseudoGreen}
\end{equation}
A similar expansion holds for $\widehat{P}({\bf r}_T,s|S_a({\bf r}_T))$.
From (\ref{laplace}), we then get

\begin{equation}\label{Tm2}
\Tm(S_a({\bf r}_T)|{\bf r}_S) P_{\rm stat}({\bf r}_T)  =  H({\bf r}_T|S_a({\bf r}_T))-H({\bf r}_T|{\bf r}_S),
\end{equation}
which  is an extension of a similar form given in \cite{Noh,nature2007}.
We then consider the large volume limit of equation (\ref{Tm2}) and define  the function  $ \phi_a({\bf r}_T|{\bf r}_S)$ as \begin{equation}\label{Tm}
  \lim_{V\to\infty} \Tm(S_a({\bf r}_T)|{\bf r}_S) P_{\rm stat}({\bf r}_T)=\phi_a({\bf r}_T|{\bf r}_S).
 \end{equation}
 As can be checked directly from definition (\ref{pseudoGreen}),  $ \phi_a({\bf r}_T|{\bf r}_S)$ satisfies the following boundary value problem in the {\it infinite} space:
\begin{equation}\label{smo}
\left\{
\begin{array}{ll}
 & \Delta_{{\bf r}} \phi_a({\bf r}_T|{\bf r})=0 \ {\rm for }\ {\bf r}\in \mathbb{R}^d\!\setminus\! B_a({\bf r}_T)\\
  & \phi_a({\bf r}_T|{\bf r})=0 \ {\rm for }\ {\bf r}\in S_a({\bf r}_T)\\
  & \displaystyle \int_{{\bf r}\in S_a({\bf r}_T)}\!\!\!\!\!\!\partial_n \phi_a({\bf r}_T|{\bf r})dS({\bf r})=1.
\end{array}\right.
\end{equation}
Equations (\ref{Tm},\ref{smo}) completely define the large volume asymptotics of the FPT and constitute the central result of our method.
Indeed, rephrased as above, the problem amounts to determining an electrostatic potential outside a conducting surface $S_a({\bf r}_T)$ of charge unity, which has been extensively studied. When  $S_a({\bf r}_T)$ is a sphere the solution is straightforward and  yields for the mean FPT:
\begin{equation}
\lim_{V\to\infty}\Tm/V=\left\{
\begin{array}{ll}
\displaystyle \frac{1}{2\pi}\ln(r/a) & \; {\rm for}\; d=2\\
\displaystyle \frac{\Gamma(d/2)}{2\pi^{d/2}}\left(\frac{1}{a^{d-2}}-\frac{1}{r^{d-2}}\right) & \; {\rm for}\; d\ge 3
\end{array}\right.
\end{equation}
where $r=|{\bf r}_T-{\bf r}_S|$. This result  is compatible with the form found in \cite{Condamin2005a,Condamin2007} using a different method. Besides the very useful  analogy with potential theory, we will show in the following that the advantage of the formulation (\ref{Tm},\ref{smo}) is threefold.
(i) First, it can be  adapted to other geometries and in particular to various examples of extended targets, such as an escape window in the domain boundary. It therefore extends the main result  of \cite{nature2007} obtained in discrete space for a point-like target.
(ii) Second, as the derivation of equations (\ref{Tm},\ref{smo}) is independent of the operator $\Delta$, it can be reproduced for any  displacement operator ${\cal L}$. In the general case, equations (\ref{Tm},\ref{smo}) still hold, but with $\Delta$ to be substituted by the adjoint operator ${\cal L}^+$, and $\partial_n \phi_a$ to be substituted by the flux of $\phi_a$. (iii) Last, and following the previous remark, the formulation (\ref{Tm},\ref{smo}) can be extended to the case of a random walker experiencing an external force field.

We first show that equations (\ref{Tm},\ref{smo}) can be extended to the example of an escape window $W_a({\bf r}_T)$ of  small typical radius $a$ centered at ${\bf r}_T\in \partial\Omega $, which is precisely the narrow escape problem. It will be useful to write $W_a({\bf r}_T)=\partial\Omega\cap B_a^\epsilon({\bf r}_T) $ where $B_a^\epsilon({\bf r}_T)$ is a small volume of typical thickness $\epsilon$. We then set  $S_a^\epsilon({\bf r}_T)\equiv \Omega\cap \partial B_a^\epsilon({\bf r}_T)$. We now derive the mean NET through $S_a^\epsilon({\bf r}_T)$, and we will use the fact that $\lim_{\epsilon\to0} S_a^\epsilon({\bf r}_T)=W_a({\bf r}_T)$ to obtain the mean NET through $W_a({\bf r}_T)$. Following step by step the previous derivation, we obtain:
\begin{equation}\label{Tm3}
\Tm(S_a^\epsilon({\bf r}_T)|{\bf r}_S) P_{\rm stat}({\bf r}_T)  =  H({\bf r}_T|S_a^\epsilon({\bf r}_T))-H({\bf r}_T|{\bf r}_S)
\end{equation}
 We now take the infinite volume limit keeping $ {\bf r}_T-{\bf r}_S$ fixed, with the prescription that $\Omega$ tends to  the half space $ \mathbb{R}_+^d$ delimited by the hyperplane containing $W_a({\bf r}_T)$, and define:
\begin{equation}\label{Tm4}
  \lim_{V\to\infty} \Tm(S_a^\epsilon({\bf r}_T)|{\bf r}_S) P_{\rm stat}({\bf r}_T)=\phi_a^\epsilon({\bf r}_T|{\bf r}_S).
\end{equation}
One can show that $\phi_a^\epsilon({\bf r}_T|{\bf r})$ then satisfies:

\begin{equation}\label{smo2}
\left\{
\begin{array}{ll}
 & \Delta_{{\bf r}} \phi_a^\epsilon({\bf r}_T|{\bf r})=0 \ {\rm for }\ {\bf r}\in \mathbb{R}_+^d\!\setminus\!B_a^\epsilon({\bf r}_T)\\
  & \phi_a^\epsilon({\bf r}_T|{\bf r})=0 \ {\rm for }\ {\bf r}\in S_a^\epsilon({\bf r}_T)\\
 & \displaystyle \partial_{n} \phi_a^\epsilon({\bf r}_T|{\bf r})=0 \ {\rm for }\ {\bf r}\in \partial\mathbb{R}_+^d\\
  & \displaystyle \int_{{\bf r}\in S_a^\epsilon({\bf r}_T)}\!\!\!\!\!\!\partial_n \phi_a^\epsilon({\bf r}_T|{\bf r})dS({\bf r})=1.
\end{array}\right.
\end{equation}
This shows that $\phi_a^\epsilon({\bf r}_T|{\bf r})$ is the electrostatic potential in a half space delimited by an isolating hyperplane containing a conducting window $S_a^\epsilon({\bf r}_T)$ of charge unity. Taking $\epsilon$ to 0 gives the mean NET through
$W_a({\bf r}_T)$. In the case of a spherical window  $W_a({\bf r}_T)$,   the solution of  (\ref{smo2}) can be exactly given. For $d=3$, we obtain    in oblate spheroidal coordinates \cite{hill}:
\begin{equation}\label{net3}
\lim_{V\to\infty}\Tm/V= \frac{1}{2\pi a}\arctan\xi=\frac{1}{4a}-\frac{1}{2\pi r}+o\left(\frac{1}{r}\right)
\end{equation}
where $\xi$ depends on  cartesian coordinates according to
\begin{equation}
\frac{z^2}{a^2\xi^2}+\frac{x^2+y^2}{a^2(\xi^2+1)}=1.
\end{equation}
For $d=2$, we use elliptic coordinates and get
\begin{equation}\label{net2}
\lim_{V\to\infty}\Tm/V= \frac{\mu}{\pi }\sim_{r\to\infty}\frac{1}{\pi}\ln(r/a)
\end{equation}
where $\mu$ depends on  cartesian coordinates according to
\begin{equation}
\frac{x^2}{a^2\cosh^2\mu}+\frac{y^2}{a^2\sinh^2\mu}=1.
\end{equation}
We stress that expressions (\ref{net3},\ref{net2}) of the mean NET are exact for any  position of the source point $\rr_S$. They therefore extend the results of \cite{Grigoriev02,Singer06,Schuss2007}, which give the same small $a$ limit, but where  the dependence on the source position was not given.

We now generalize equations (\ref{Tm},\ref{smo}) to the case of a generic displacement operator ${\cal L}$ such that the  stationary distribution is \emph{uniform} $P_{\rm stat}=1/V$, which  actually underlies many models of transport in complex media \cite{Bouchaud1990}. We here assume that  $S_a({\bf r}_T)$ is a sphere, and following \cite{BenAvraham,nature2007},
we further assume that the infinite space propagator $P_0 $ satisfies the standard scaling:
\begin{equation}
P_0({\bf r},t|{\bf r}') \sim t^{-d_f/d_w} \Pi\left(
\frac{|{\bf r}-{\bf r}'|}{t^{1/d_w}}\right),
\label{scaling}
\end{equation}
 where the fractal dimension $d_f$ characterizes the
accessible volume $V_r \sim r^d_f$ within a sphere of radius $r$,   and  the walk dimension $d_w$  characterizes  the distance $r \sim t^{1/d_w}$ covered by a random walker
in a given  time $t$. This formalism in particular covers the case of a random walk on a random fractal like critical percolation clusters, which gives a representative example  of subdiffusive behavior due to crowding effects \cite{BenAvraham} and could mimic in a first approximation the cellular environment. Note that we here implicitly require that the trajectories and the medium have length scale invariant properties which ensure the existence of $d_w$ and $d_f$.
Substituting the scaling (\ref{scaling}) in the definition (\ref{pseudoGreen}), we obtain from (\ref{Tm}) the large $V$ equivalence:
\begin{equation}
\lim_{V\to\infty}\Tm/V = \left\{
\begin{array}{ll}
\alpha (a^{d_w-d_f} - r^{d_w-d_f}) & \; {\rm for}\; d_w<d_f\\
\alpha \ln (r/a) & \; {\rm for}\; d_w=d_f\\
\alpha ( r^{d_w-d_f}-a^{d_w-d_f}) & \; {\rm for}\; d_w>d_f
\end{array}
\right.
\label{scalingt}
\end{equation}
Strikingly, the constant $\alpha$ does not depend on the confining
domain but only on the scaling function $\Pi$:
\begin{equation}
\alpha = \left\{
\begin{array}{ll}
\displaystyle \int_0^\infty \frac{\Pi(u^{-1/{d_w}})}{u^{d_f/d_w}}du & \; {\rm for}\; d_w<d_f\\
\displaystyle   d_w\Pi(0) & \; {\rm for}\; d_w=d_f\\
\displaystyle \int_0^\infty \frac{du}{u^{d_f/d_w}}\left(\Pi(0)-\Pi(u^{-1/d_w})\right) & \; {\rm for}\; d_w>d_f
\end{array}
\right.
\label{alpha}
\end{equation}
Expressions (\ref{scalingt}) therefore explicitly elucidate the dependence of the mean FPT on the geometrical parameters $V$ and $r$. Note that for a generic target surface $S_a({\bf r}_T)$, the $r$ dependence in (\ref{scalingt}) still holds for $r\gg a$, but  the $a$ dependence is changed by a  numerical factor. Indeed, the $r$ dependence is fully determined by writing the conservation of the flux, which does not depend on the window shape at large $r$.
As previously equations (\ref{scalingt}) permit to obtain the mean NET: if we assume that the exit window $S_a^\epsilon({\bf r}_T)$ is a half sphere of radius $a$, the mean NET will be exactly given by two times the mean FPT (\ref{scalingt}). For a generic window, and in particular in the case of a disk, the $r$ dependence is unchanged for $r\gg a$, but  the $a$ dependence is modified by a  numerical factor.

Remarkably, equation (\ref{scalingt})  highlights two regimes. When the exploration is not compact ($d_w<d_f$), as in the case of a brownian particle in the 3--dimensional  space, the dependence on the starting point disappears at large $r$. On the other hand, in the case of compact exploration ($d_w\le d_f$), as for 2--dimensional diffusion or subdiffusion on fractals, the mean NET diverges at large $r$ and the starting point position is crucial.

Last, we consider the case of a brownian particle in the presence of a force field $F(\rr)=-\nabla_\rr \Phi(\rr)$. In the context of biological cells, such force can be induced by the coupling of the particle to molecular motors which perform  a directional motion along cytoskeletal filaments \cite{alberts}.  We assume that the target is a sphere $S_a({\bf r}_T)$, and that the force field is spherically symmetric and centered at $\rr_T$, situation which mimics asters of cytoskeletal filaments, which are ubiquitous in cells \cite{alberts}.  We set the gauge such that $\Phi(\rr)=0$ for $\rr\in B_a({\bf r}_T)$. Equations (\ref{Tm},\ref{smo}) then hold with $\Delta_\rr$ to be replaced by the adjoint operator ${\cal L}^+$ governing the evolution of the propagator \cite{Kampen1992}:
\begin{equation}
 {\cal L}^+=F(\rr)\nabla_\rr+\Delta_\rr.
\end{equation}
We then solve ${\cal L}^+ \phi_a({\bf r}_T|{\bf r})=0$ with the same boundary conditions as in (\ref{smo}), and
write the stationary distribution
\begin{equation}
 \displaystyle P_{\rm stat}(\rr)=\frac{e^{-\Phi(\rr)}}{\int_{\Omega}\! d\Omega(\rr')\; e^{-\Phi(\rr')}}.
\end{equation}
Using equation (\ref{Tm}), we finally  get the large $V$ equivalence of the mean FPT:
\begin{equation}\label{sink}
 \Tm\sim\displaystyle \frac{\Gamma(d/2)}{2\pi^{d/2}} \left(\int_{\Omega}\!\!\! d\Omega(\rr')\; e^{-\Phi(\rr')}\right)\left(\int_a^r\!\!\! u^{1-d} e^{\Phi(u)}\right).
\end{equation}
As previously, the mean NET through a spherical window is obtained as two times the mean FPT. One should note that the  volume dependence entirely lies in the first integral factor of (\ref{sink}), which is suitable for a quantitative analysis. On the other hand,  the $r$ dependence is fully contained in the second integral factor and will depend on the specific shape of $\Phi$. The volume dependence of (\ref{sink})  agrees with the one found in \cite{Singer2006}, where however the $r$ dependence was not given.

We now give the explicit example of a divergence less force, which is a first approximation of the force experienced by a particle in an aster of cytoskeletal filaments, if one assumes that the force is proportional to the local concentration of filaments. This situation also describes a brownian particle advected in an incompressible hydrodynamic flow.  For $d=3$ such force is given by $F(r)=-\gamma/r^2$ and the equivalent potential governing the dynamics can be taken as $\Phi(\rr)=\gamma(1/r-1/a)$. Applying (\ref{sink}), we get
\begin{equation}\label{force3}
\lim_{V\to\infty}\Tm/V=\frac{1}{4\pi\gamma}\left(e^{\gamma(1/a-1/r)}-1\right)
\end{equation}
for a generic domain shape. Note that the $r$ dependence is modified by the hydrodynamic flow, while the $V$ dependence at large $V$ is unchanged for any flow intensity $\gamma$.
For $d=2$ the force is given by $F(r)=-\gamma/r$ and the equivalent potential can be taken as $\Phi(\rr)=\gamma\ln(r/a)$. Applying (\ref{sink}) in the case of a  domain whose boundary is parameterized by $R(\theta)$ in polar coordinates, we get the large volume equivalence
\begin{equation}\label{force2}
\Tm\sim\frac{1}{\gamma(2-\gamma)}\left(\int_0^{2\pi}\!\!\frac{d\theta}{2\pi} R^{2-\gamma}(\theta)-\frac{\gamma}{2}a^{2-\gamma}\right)\left(r^\gamma-a^\gamma\right).
\end{equation}
As opposed to the $d=3$ case, the $V$ dependence is now modified by the force.  Interestingly we find a transition for $\gamma=2$. For $\gamma<2$, the mean FPT will scale like $V^{1-\gamma/2}$ at large $V$, while the $V$ dependence disappears for $\gamma>2$.
In the case of a target centered in a spherical domain the results of \cite{RednerBook} can be straightforwardly recovered from (\ref{sink}), and are compatible with (\ref{force3},\ref{force2}).

To conclude, we have proposed a general theory  which  provides explicitly the scaling dependence of the mean NET on both the volume of the confining domain and the source-aperture distance for a wide range of transport processes, including anomalous diffusion and transport in the presence of a force field, which are relevant to biological situations. In particular, we find that the dependence of the mean NET on  the source aperture distance is crucial when the exploration is compact,  as it is  the case for the subdiffusive behavior in crowded environment which is observed in cells. Our model also permits to take into account the active transport due to molecular motors.


\end{document}